\documentstyle[twocolumn,prl,floats,aps,epsfig]{revtex}

\begin{document}

\draft

\wideabs{
\title{Normal-state resistivity anisotropy in underdoped
RBa$_2$Cu$_3$O$_{6+x}$ crystals}

\author{A. N. Lavrov\cite{ANL}, M. Yu. Kameneva, L. P. Kozeeva}

\address{Institute of Inorganic Chemistry, Siberian Branch of RAS,
Lavrentyeva-3, 630090 Novosibirsk, Russia}

\date{\today}

\maketitle

\begin{abstract}
We have revealed new features in the out-of-plane resistivity $\rho_c$ of
heavily underdoped RBa$_2$Cu$_3$O$_{6+x}$ (R=Tm,Lu) single crystals, which
give evidence for two distinct mechanisms contributing the $c$-axis
transport. We have observed a crossover towards "metal-like" ($\partial
\rho_c / \partial T > 0$) behavior at the temperature $T_m$ which quickly
{\it increases} with decreasing doping. The "metal-like" conductivity
contribution dominates at $T < T_m$ and provides a saturation of the
resistivity anisotropy, $\rho_c / \rho_{ab}$. The antiferromagnetic
ordering is found to block this "metal-like" part of the $c$-axis
conductivity and complete decoupling of CuO$_2$ planes, which may be the
reason of superconductivity disappearance.
\end{abstract}

}

The primary indication of an unusual normal state in high-$T_c$ cuprates is
the contrasting behavior of the in-plane ($\rho_{ab}$) and out-of-plane
($\rho_c$) resistivity. In most cuprates, for instance
La$_{2-x}$Sr$_x$CuO$_4$, underdoped YBa$_2$Cu$_3$O$_{6+x}$,
Bi$_2$Sr$_2$CuO$_y$, the metal-like electron transport along CuO$_2$ planes
coexists with a non-metallic conductivity between planes, and the
resistivity anisotropy, $\rho_c/\rho_{ab}$, diverges with decreasing
temperature till the superconducting transition interrupts this tendency
\cite{LaSr1,ybco1,Ando1}. This behavior violating the conventional concept
of band electron transport has brought into being many theories which imply
blocking the $c$-axis coherent transport and charge confinement within the
CuO$_2$ planes \cite{Ander,Kumar,coher}. The salient consequence of charge
confinement is a possibility of superconductivity owing to interlayer pair
tunneling \cite{Ander,Kumar}. The two-dimensional behavior is considered
thus as a key quality of that unusual normal state giving rise to
high-$T_c$ superconductivity.

Important exceptions from this straightforward picture were however found,
and the best known one is YBa$_2$Cu$_3$O$_7$ (Y-123) which is a 90-K
superconductor, but possesses a metallic out-of-plane conductivity
\cite{Ito}. A crossover towards the coherent $c$-axis electron transport
with decreasing temperature was recently found in YBa$_2$Cu$_4$O$_8$
(Y-124) \cite{cros1,cros2}. This peculiar behavior of Y-123 and Y-124
systems was attributed to the metallic conductivity of their Cu-O chains.
In contrast to expectations, a temperature crossover in $\rho_c(T)$
resembling that in Y-124 was observed also in heavily underdoped
RBa$_2$Cu$_3$O$_{6+x}$ (R=Y, rare earth) crystals in which, obviously, the
Cu-O chains were destroyed \cite{Lavr1}.

Analyzing experimental data for highly anisotropic high-$T_c$ cuprates, one
should take into account crystal perfection problems. Stacking faults can
well block the $c$-axis conductivity and give rise to insulating
$\rho_c(T)$, while an apparent metallic behavior and crossovers can
originate from numerous screw dislocations \cite{screw} short-circuiting
the whole set of CuO$_2$ planes. Recently \cite{whisk} we have found that
R-123 crystals can grow not only as conventional thin or thick plates, but
also like whiskers along the $b$ axis. We succeeded in growing whisker-like
Tm-123 crystals which had a shape of thin, wide bars with the shiny
$bc$-faces being the {\it largest} ones. These unique crystals are very
attractive for studying the resistivity anisotropy $\rho_c(T)/
\rho_{ab}(T)$. While their shape is suitable for measuring {\it both}
resistivity components, the growth mechanism being distinct from that of
platelets implies the absence of screw dislocations along the $c$-axis,
i.e. in the direction transverse to the crystal growth one.

In the present work, using mainly these whisker-like crystals, we
demonstrate that the out-of-plane conductivity in RBa$_2$Cu$_3$O$_{6+x}$
{\it inherently} contains two distinct contributions associated presumably
with two types of charge carriers. The first contribution is
temperature-activated and provides the familiar contrast between
$\rho_c(T)$ and $\rho_{ab}(T)$. The second one roughly follows the in-plane
conductivity $\sigma_{ab}$ though reduced by 4 orders of magnitude. This
contribution dominates the low-temperature $c$-axis transport, induces a
crossover in $\rho_c(T)$ and prevents the resistivity anisotropy from
diverging at low $T$. In contrast to the Y-124 system \cite{cros2} this
metal-like conduction cannot be associated with Cu-O chains which are
destroyed in our underdoped crystals.

\begin{figure}[t]
\begin{center}
\leavevmode
\leftskip-15pt
\epsfxsize=1.2\columnwidth
\epsffile{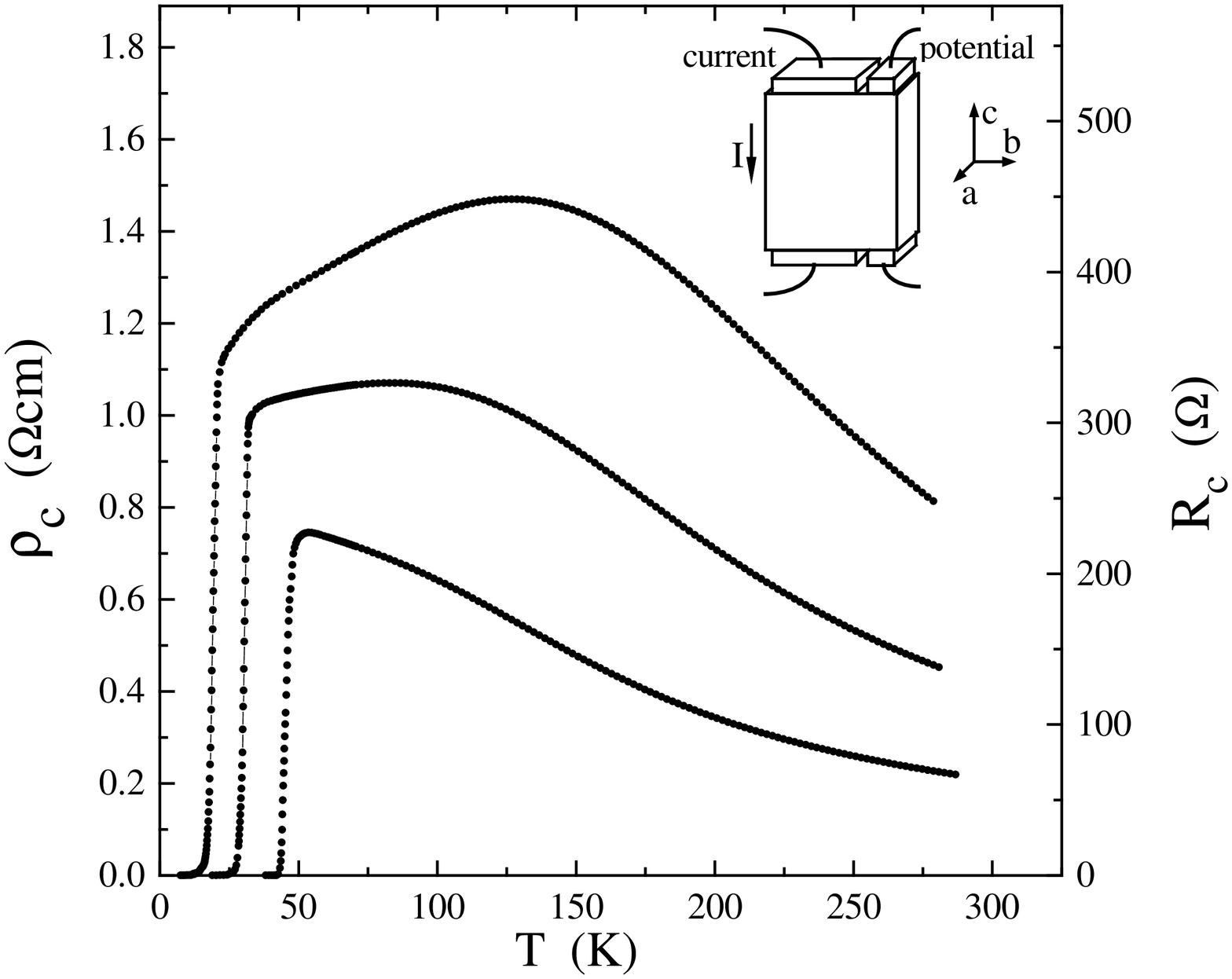}
\caption{The out-of-plane resistivity, $\rho_c(T)$, of the
TmBa$_2$Cu$_3$O$_{6+x}$ "whisker-like" single crystal at various
oxygen contents, $x\approx 0.41, 0.44, 0.49$ (top to bottom). Inset:
the contact configuration.}
\label{fig1}
\end{center}
\end{figure}

Both the plate- and whisker-like (Tm,Lu)Ba$_2$Cu$_3$O$_{6+x}$ crystals were
grown by the flux method \cite{whisk} and their oxygen stoichiometry was
varied by subsequent high-temperature annealing \cite{Lavr1}. Measurements
of $\rho_c(T)$ and $\rho_{ab}(T)$ were performed by the four-probe method
on two samples cut always from the same single crystal. Further description
will concentrate mainly on whisker-like Tm-123 crystals, since their shape
and absence of screw $c$-axis dislocations allow a straightforward analysis
of $\rho_c$ data. For example, the sample in Fig.~\ref{fig1} was
$45\times400\times550\,\mu$m$^3$ with the largest dimension along the
$c$-axis. Mapping this crystal on an isotropic model, one obtains a thin,
long wire ($\approx 0.01\times0.1\times10$~mm$^3$), for which evaluating
$\rho_c$ from raw data holds, obviously, no problems. For $\rho_{ab}$
measurements, in their turn, a narrow bar of $45\times800
\times140\,\mu$m$^3$ was cut from the same whisker-like crystal.

In the RBa$_2$Cu$_3$O$_{6+x}$ system, superconductivity mostly hides the
low-$T$ region and the normal-state resistivity can be measured down to
fairly low temperatures only in a narrow doping range in the vicinity of
the AF-SC phase boundary. A selection of $\rho_c(T)$ curves obtained within
this heavily underdoped region is presented in Fig.~\ref{fig1}. In contrast
to what one could expect the $c$-axis resistivity does not grow sharply
with decreasing temperature, but passes through a maximum at $T_m$ and
begins to drop. For the sample with $x\approx 0.41$, $\rho_c$ monotonically
decreases within a wide temperature range from 127 K down to $T_c \approx
19$~K. A tendency of $\rho_c$ to saturate is apparent for $x\approx0.49$ as
well, but for such and higher doping levels the $\rho_c$ crossover is
masked, because $T_m$ quickly decreases with doping, while
superconductivity in its turn hides the larger temperature range. This is
why other studies of YBa$_2$Cu$_3$O$_{6+x}$ dealing mainly with $x\geq 0.6$
\cite{ybco1} could not reveal this $\rho_c$ peculiarity. The crossover
observed gives evidence for the change of the dominating conductivity
mechanism, and implies most likely that the out-of-plane conductivity
contains two contributions, the balance of which determines the shape of
$\rho_c(T)$ curves.

\begin{figure}[t]
\begin{center}
\leavevmode
\leftskip-10pt
\epsfxsize=1.25\columnwidth
\epsffile{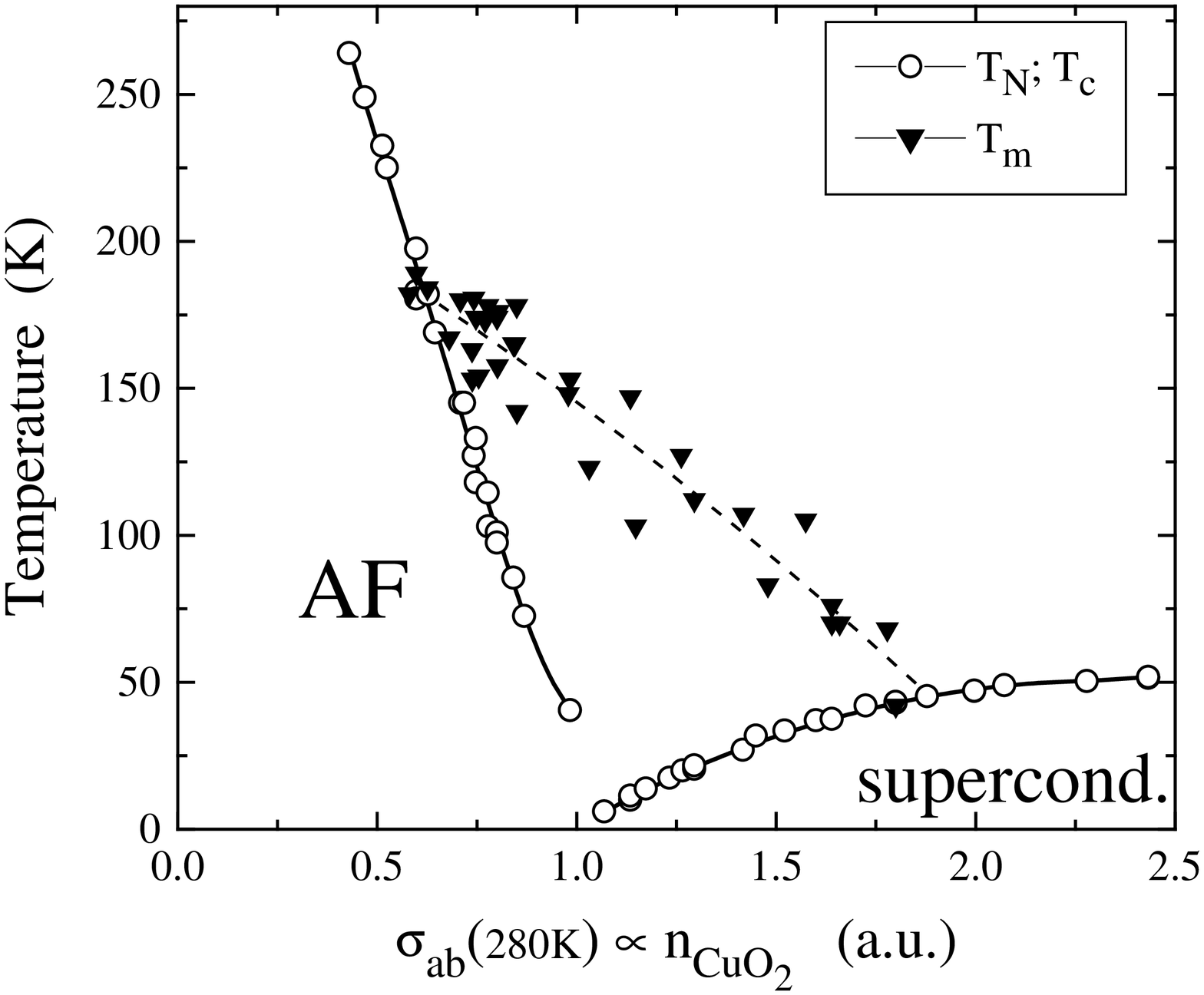}
\caption{A cumulative diagram for TmBa$_2$Cu$_3$O$_{6+x}$ and
LuBa$_2$Cu$_3$O$_{6+x}$ crystals. The crossover temperature $T_m$
together with the AF and SC transition temperatures, $T_N$, $T_c$, are
presented as a function of the in-plane conductivity $\sigma_{ab}$.}
\label{fig2}
\end{center}
\end{figure}

The crossover temperatures $T_m$ determined from 34 resistivity curves
measured on 9 plate- and whisker-like Tm-123 and Lu-123 crystals with
different oxygen contents are collected in Fig.~\ref{fig2}. The data are
plotted on the phase diagram in which both the Neel temperature $T_N$ and
the superconducting transition temperature $T_c$ are determined from
resistivity measurements \cite{Lavr1} and presented as a function of the
in-plane conductivity $\sigma_{ab}$, which is roughly proportional to the
hole density in the CuO$_2$ planes. This diagram presentation looks similar
to the usual $T-x$ one, but qualitatively accounts for the influence of
oxygen ordering on the hole density \cite{Lavr1}. As can be seen $T_m$
quickly decreases with increasing doping (increasing $x$) and somewhere at
the 60-K plateau the crossover line gets under the SC region. The Cu-O
chains, when perfectly ordered, possess the metallic conductivity
\cite{Ito,cros2}, and one could expect the range of metallic behavior to
extend with increasing oxygen content. Obviously the phenomenon we are
dealing with is of a different nature. The metal-like conductivity
component dominates just in the region where the Cu-O chains are destroyed.
The heavily underdoped crystals have no other conducting subsystem besides
CuO$_2$ planes and one hence has no choice but to attribute the metal-like
conduction to the direct interplane charge transport.

Each conductivity contribution can be analyzed separately at a distance
from the crossover line, i.e. in the range where it dominates. The
metal-like contribution roughly tracks the in-plane conductivity behavior,
while the activated one can be fitted by exponential expressions, the
simplest of which is that of variable range hopping. The crossover behavior
can be thus described as $\sigma_c(T) = K\sigma_{ab}(T) + C\exp(-B/T^{1/
4})$, with the conductivity anisotropy approaching a constant value at low
$T$.

To obtain additional information on the conductivity contributions we had
available two simple approaches. Firstly, we could move the crystal from
the SC to AF doping region, see Fig.~\ref{fig2}, and analyze how the
long-range magnetic order influences $\rho_c$. To test the nature of the
metal-like conductivity one has to place the Neel temperature in that
temperature region where this contribution dominates, i.e. $T_N$ should be
$<100$~K. The second possibility is to use the well studied phenomenon of
chain-layer oxygen ordering, see Ref.~\cite{Lavr1} and references therein,
as a convenient way of tuning the hole density in CuO$_2$ planes. The
hole-doping level can be reduced by $\approx 20\,\%$ by heating the crystal
to $\approx 120\,^o$C with subsequent quenching, and it can be gradually
restored simply by room-temperature aging. The advantage of this procedure
is that both the stoichiometry and the contact configuration remain exactly
unchanged.

\begin{figure}[t]
\begin{center}
\leavevmode
\leftskip-14pt
\epsfxsize=1.15\columnwidth
\epsffile{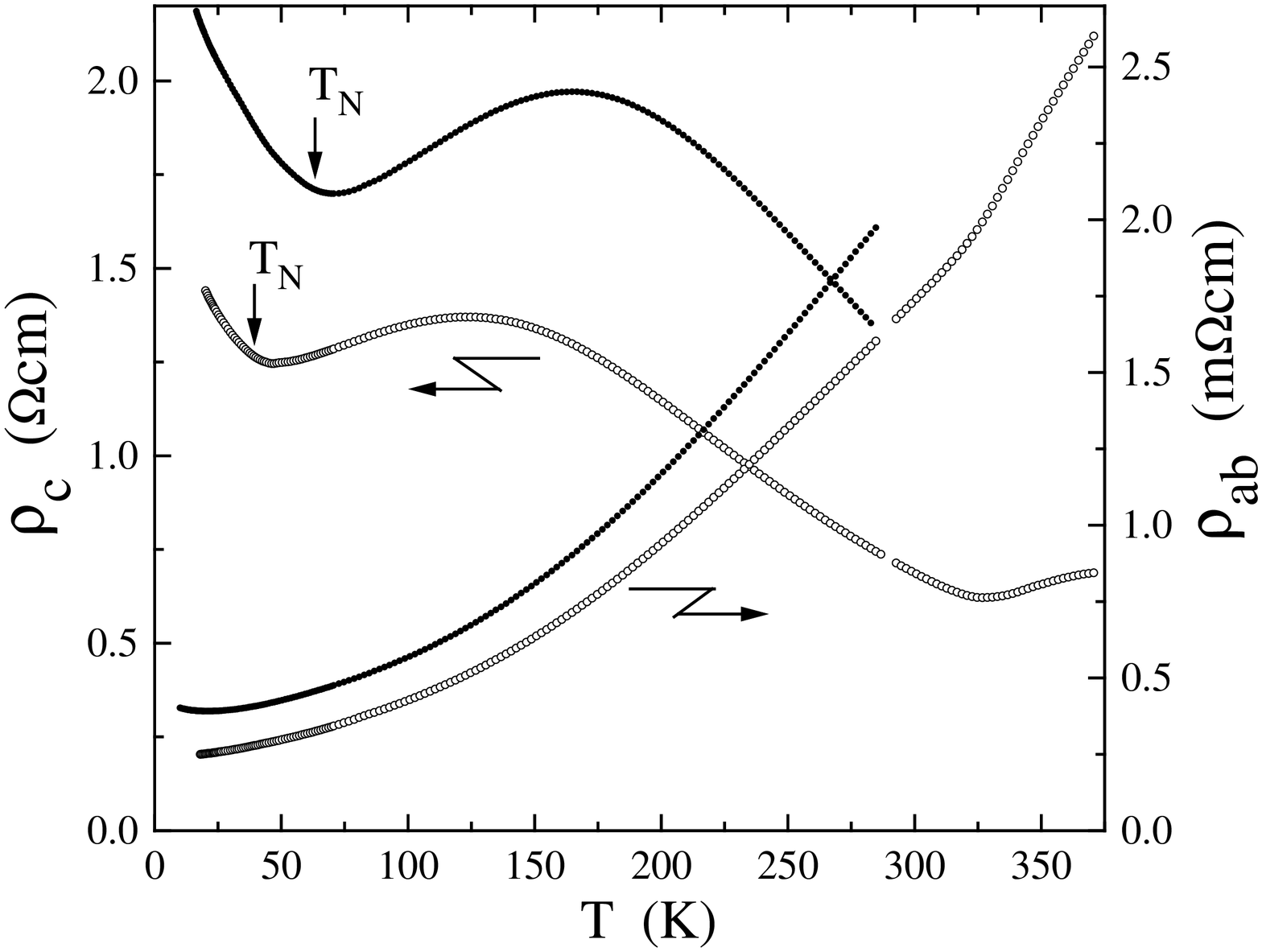}
\caption{In-plane, $\rho_{ab}(T)$, and out-of-plane, $\rho_c(T)$,
resistivity of TmBa$_2$Cu$_3$O$_{6+x}$ ($x\approx 0.37$) crystals.
Measurements were performed immediately after quenching (solid
circles) and after 5 days aging at room temperature (open circles).}
\label{fig3}
\end{center}
\end{figure}

Fig.~\ref{fig3} combines both approaches and presents the $\rho_c(T)$ and
$\rho_{ab}(T)$ data obtained for non-superconducting crystals. One can see
that at $T>20$~K \cite{lowT} $\rho_{ab}$ has almost parabolic temperature
dependence and roughly scales with hole doping. The interplane resistivity
retains the pronounced crossover, but an anomaly associated with the AF
ordering \cite{Lavr1} has appeared instead of SC transition, compare with
Fig.~\ref{fig1}. A step-like increase of $\rho_c$ occurs upon cooling below
the Neel temperature. We notice that $T_N$ is essentially below the
crossover point, and the $\rho_c$ anomaly is located in the region where
the metal-like conductivity component undoubtedly dominates, but at the
same time no peculiarity is observed on $\rho_{ab}(T)$ curves. This is
probably the most visual evidence that the $\rho_c$ crossover cannot be
associated with some admixture of the in-plane conductivity.

\begin{figure}[t]
\begin{center}
\leavevmode
\leftskip-10pt
\epsfxsize=1.3\columnwidth
\epsffile{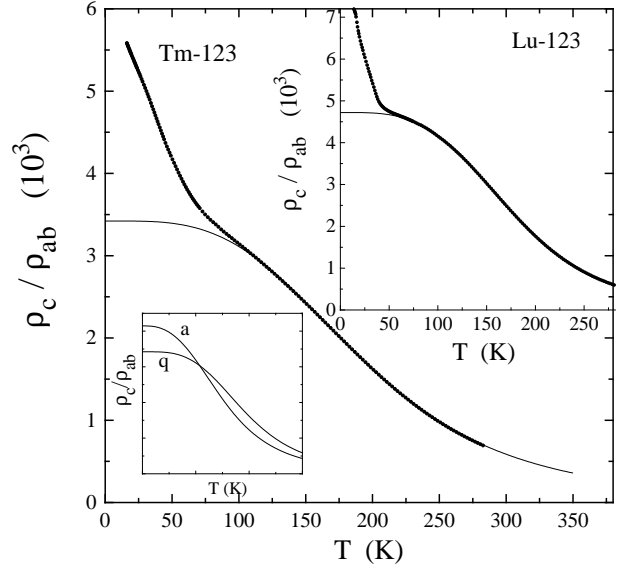}
\caption{Main graph: resistivity anisotropy $\rho_c/\rho_{ab}$ for
"whisker-like" TmBa$_2$Cu$_3$O$_{6.37}$, solid line - a fit of the
regular temperature dependence ($T > T_N$). Left insert: fits of the
resistivity anisotropy for quenched and aged states of the same
crystal. Right inset: resistivity anisotropy for a "plate-like"
LuBa$_2$Cu$_3$O$_{6.34}$ crystal.}
\label{fig4}
\end{center}
\end{figure}

Fig.~\ref{fig4} presents the $(\rho_c/\rho_{ab})(T)$ curve obtained from
the data shown by solid circles in Fig.~\ref{fig3}. Above $T_N$ the
resistivity anisotropy can be well fitted by the empirical expression
implying that $\sigma_c$ contains two contributions. One can see that in
the absence of antiferromagnetic ordering, the anisotropy would saturate at
a value of several thousand. The long-range AF ordering obviously blocks
the metal-like conductivity contribution and completes decoupling of
CuO$_2$ planes. For comparison, in the right inset we show the anisotropy
data obtained for a highly homogeneous plate-like Lu-123 crystal. The only
apparent difference is a sharper AF transition, which makes the regular
behavior of the resistivity anisotropy and, particularly, its tendency to
saturation more spectacular. The same anisotropy saturation is obviously
characteristic of SC samples as well, since $\rho_c(T)$ curves for crystals
with $T_c \approx 19$~K and $T_N \approx 65$~K,
Figs.~\ref{fig1},\ref{fig3}, not much differ above $T_c$($T_N$).

The left inset in Fig.~\ref{fig4} compares fits of regular ($T > T_N$)
$(\rho_c/\rho_{ab})(T)$ dependences for two hole-doping levels of the
same crystal, and illustrates one more surprising result. The
resistivity anisotropy at high $T$ considerably decreases with
increasing density of carriers, but its extrapolated low-temperature
value, in contrast to what one could expect, {\it increases} with
increasing doping. The metal-like conductivity contribution under
discussion has therefore nothing to do with the metallic conductivity
in fully oxygenated YBa$_2$Cu$_3$O$_7$. Instead, it is the
temperature-activated contribution which should gradually acquire
metallic features with $x\rightarrow 1$ and begin to dominate over the
whole temperature range.

Few theories consider a possibility of crossover towards metal-like
$c$-axis conductivity at low temperatures in underdoped cuprates
\cite{coher,polar}. The general problem which emerges upon constructing a
picture based on multi-band conduction, Kondo-scattering, etc., is that
$\rho_c$ at any temperature appears to be too large to fit the concept of
coherent conduction. To ascribe the metal-like conductivity observed to the
coherent transport one has either to suppose that only a small fraction of
carriers participates in the $c$-transport, or to introduce new heavy
quasiparticles \cite{polar}. On the other hand, the incoherent $c$-axis
conductivity can also track the behavior of $\sigma_{ab}$, if just the
strong in-plane scattering blocks the interplane transitions \cite{Kumar}.

The most attractive approach for the temperature-activated conductivity
contribution is to attribute it to Cu-O chains. Really, if perfect Cu-O
chains in Y-123 and Y-124 systems possess the metallic conductivity
\cite{Ito,cros2}, fragmented ones in underdoped crystals should naturally
provide the hopping electron transport. This assumption would explain the
observed strong dependence of $\rho_c(T)$ on both the oxygen content and
oxygen ordering.

However, we probably should search for a more general explanation for the
puzzling $c$-transport rather than that based on structural peculiarities
of R-123. Actually, apart from the resistivity scale, similar behavior
($\rho_{ab} \propto T^2$ and a crossover in $\rho_c$ at $\approx 120$~K)
was observed in non-cuprate layered system Sr$_2$RuO$_4$ \cite{SrRuO}, and
the anisotropy saturation at low $T$ was reported for
La$_{2-x}$Sr$_x$CuO$_4$ \cite{Ando1}. Of course, approaches not assuming
any role of Cu-O chains could be suggested. The concept of charge
confinement, for instance, implies blocking of the interplane
single-particle tunneling, but retains a possibility of coherent transport
for pairs \cite{Ander,Kumar}. The two types of carriers, namely bosons and
thermally excited fermions, could thus be responsible for distinct
conductivity contributions observed. The pair formation not necessarily
results in superconductivity and these are preformed pairs which are often
considered as a cause of pseudogap effects in cuprates \cite{polar,pairs}.
Just a glance is enough to find similarity between the crossover
temperature $T_m$ in Fig.~\ref{fig2} and the pseudogap crossover
temperature $T^*$ in popular phase diagrams suggested for cuprates
\cite{gap}. Some difference existing in the scales, but not in the doping
dependences, is not valuable, since both $T_m$ and $T^*$ correspond to
arbitrary determined crossover points.

The blocking phenomenon due to AF ordering holds a problem for all
cited above models and remains yet to be explained. If $\rho_c$ was
controlled by interplane scattering \cite{Kumar}, it would not change
at $T_N$, since the in-plane scattering obviously does not undergo
considerable variation. On the other hand, if the metal-like
conductivity component originates from preformed pairs
\cite{Ander,Kumar,polar}, one necessarily faces the question how
spinless carriers interact with the magnetic order. An interesting
consequence within the preformed-pair model is that blocking of
interplane pair transitions could explain why antiferromagnetism and
superconductivity hardly coexist in cuprates.

In summary, we have found new features in the out-of-plane electron
transport of heavily underdoped RBa$_2$Cu$_3$O$_{6+x}$ single
crystals. We have shown that the $c$-axis conductivity intrinsically
contains two contributions. The first one is the familiar
semiconductor-like conductivity usually observed in moderately
underdoped samples. The other looks metal-like and dominates the
interplane transport at low temperatures and low doping, where the
Cu-O chains are destroyed. Because of this contribution the
resistivity anisotropy saturates at low $T$ instead of diverging. The
finding possibly having implication for the nature or high-$T_c$
superconductivity is that in non-superconducting samples the
metal-like part of $\sigma_c$ is blocked by antiferromagnetic
ordering.

\acknowledgments ANL is grateful to Prof. V. F. Gantmakher for many
fruitful discussions.


\begin{thebibliography}{99}

\bibitem[*]{ANL} Present address: Central Research Inst. of
Electric Power Industry, 2-11-1 Iwado-kita, Komae, Tokyo 201, Japan.

\bibitem{LaSr1} Y. Nakamura and S. Uchida, Phys. Rev. B {\bf 47},
8369 (1993).

\bibitem{ybco1} K. Takenaka, K. Mizuhashi, H. Takagi, and S. Uchida,
Phys. Rev. B {\bf 50}, 6534 (1994).

\bibitem{Ando1} Y. Ando {\it et al.}, Phys. Rev. Lett. {\bf 75}, 4662
(1995); {\bf 77}, 2065 (1996).

\bibitem{Ander} S. Chakravarty, A. Sudbo, P. W. Anderson, and S. Strong,
Science {\bf 261}, 337 (1993); P. W. Anderson, cond-mat/9801267.

\bibitem{Kumar} N. Kumar, T. P. Pareek and A. M. Jayannavar, Phys. Rev. B
{\bf 57}, 13399 (1998).

\bibitem{coher} H. C. Lee and P. B. Wiegmann, Phys. Rev. B {\bf 53},
11817 (1996).

\bibitem{Ito}   T. Ito {\it et al.}, Nature {\bf 350}, 596 (1991).

\bibitem{cros1} J.-S. Zhou, J. B. Goodenough, B. Dabrowski, and K. Rogacki,
Phys. Rev. Lett. {\bf 77}, 4253 (1996).

\bibitem{cros2} N. E. Hussey {\it et al.}, Phys. Rev. Lett. {\bf 80}, 2909
(1998).

\bibitem{Lavr1} A. N. Lavrov and L. P. Kozeeva, Physica C {\bf 248},
365 (1995); {\bf 253}, 313 (1995); JETP Lett. {\bf 62}, 580 (1995).

\bibitem{screw} C. T. Lin, J. Crystal Growth {\bf 143}, 110 (1994).

\bibitem{whisk} L. P. Kozeeva {\it et al.}, J. Crystal Growth, submitted.

\bibitem{lowT}  for discussion of localization effects at low temperatures
see V. F. Gantmakher {\it et al.}, JETP Lett. {\bf 65}, 870 (1997).

\bibitem{polar} A. S. Alexandrov, V. V. Kabanov and N. F. Mott, Phys. Rev.
Lett. {\bf 77}, 4796 (1996).

\bibitem{SrRuO} F. Lichtenberg, A. Catana, J. Mannhart, and D. G. Schlom,
Appl. Phys. Lett. {\bf 60}, 1138 (1992).

\bibitem{pairs} V. B. Geshkenbein, L. B. Ioffe, and A. I. Larkin,
Phys. Rev. B {\bf 55}, 3173 (1997).

\bibitem{gap}   B. Batlogg, V. J. Emery, Nature {\bf 382}, 20 (1996).

\end{thebibliography}
\end{document}